\documentclass[12pt]{iopart}
\usepackage{graphicx}
\usepackage{bm}
\usepackage{epstopdf}
\usepackage{amsfonts}
\usepackage{amssymb}
\usepackage{latexsym,enumerate}
\usepackage{epsfig}
\usepackage{color,soul}
\usepackage[colorlinks=true,linkcolor=blue,urlcolor=blue,citecolor=blue,pdfusetitle]{hyperref}

\usepackage{fancyhdr}

\newcommand{\ket}[1]{\left\vert#1\right\rangle}

\newcommand{\eqref}[1]{(\ref{#1})}

\begin{document}

\title{Global and local thermometry schemes in coupled quantum systems}
\author{Steve Campbell}
\address{Istituto Nazionale di Fisica Nucleare, Sezione di Milano, \& Dipartimento di Fisica, Universit{\`a} degli Studi di Milano, Via Celoria 16, 20133 Milan, Italy,\\ and\\}
\address{Centre for Theoretical Atomic, Molecular, and Optical Physics, School of Mathematics and Physics, Queen's University Belfast, BT7 1NN, United Kingdom}
\author{Mohammad Mehboudi}
\address{Unitat de F\'isica Te\`orica: Informaci\'o i Fen\`omens Qu\`antics, Departament de F\'isica, Universitat Aut\`onoma de Barcelona, 08193 Bellaterra (Barcelona), Spain}
\author{Gabriele De Chiara}
\address{Centre for Theoretical Atomic, Molecular, and Optical Physics, School of Mathematics and Physics, Queen's University Belfast, BT7 1NN, United Kingdom}
\author{Mauro Paternostro}
\address{Centre for Theoretical Atomic, Molecular, and Optical Physics, School of Mathematics and Physics, Queen's University Belfast, BT7 1NN, United Kingdom}
\begin{abstract}
We study the ultimate bounds on the estimation of temperature for an interacting quantum system. We consider two coupled bosonic modes that are assumed to be thermal and using quantum estimation theory establish the role the Hamiltonian parameters play in thermometry. We show that in the case of a conserved particle number the interaction between the modes leads to a decrease in the overall sensitivity to temperature, while interestingly, if particle exchange is allowed with the thermal bath the converse is true. We explain this dichotomy by examining the energy spectra. Finally, we devise experimentally implementable thermometry schemes that rely only on locally accessible information from the total system, showing that {\it almost} Heisenberg limited precision can still be achieved, and we address the (im)possibility for multiparameter estimation in the system.
\end{abstract}
\maketitle

\fancyhf{}
\rhead{\thepage}
%\tableofcontents

\section{Introduction}
\label{intro}
Effective means to measure the properties of systems is central to all aspects of modern physics. The ability to precisely determine the working parameters of a given set-up has huge practical advantages from the formulation of accurate predictions on the behaviour of the system to quantum state preparation and manipulation~\cite{0026-1394-42-3-S08,PhysRevLett.116.061102,lombardi2001time,lloydPRL,lloyd,paris}. It is thus crucial to have the most accurate characterization of the key parameters describing the system's evolution, preferably in a minimally disturbing way for the dynamics that we aim at implementing. This is even more relevant for systems of difficult direct addressability or endowed with many mutually interacting degrees of freedom. In general, the determination of the features of a given model in these contexts requires measurements that are strongly disruptive for the fragile state of the system.

In this respect, the approach based on probing quantum evolutions, where a fully controllable probe is coupled to the system of interest and subsequently measured to extract the relevant information, is very promising as it allows for the implementation of weakly disruptive strategies by means of indirect interrogation~\cite{PhysRevLett.114.220405,campbellPRA,gooldPRA,maniscalcoPRA,giorgiPRA,rossiPRA,plenioPRA,PhysRevA.91.033631}. Moreover, such approaches often require the application of sophisticated techniques for parameter estimation that aim at determining the best preparation and measurement of the probe and are explicitly designed to achieve the best possible accuracy of estimation allowed by classical and quantum mechanics. In this context, recent advances in quantum metrology have opened new exciting perspectives for determining the working parameters of broadly applicable Hamiltonians as well as pushing the achievable boundaries of thermometry~\cite{1367-2630-17-5-055020, 2015arXiv150407787D, PhysRevA.92.052112,PhysRevA.88.063609, ParisJPA}.

In this work we focus on a model of wide experimental appeal, namely two coupled bosonic modes, and we explore the potential to accurately estimate its equilibrium temperature. This model encompasses a wide variety of relevant physical settings, such as loaded double-wells~\cite{campbellSciRep}, certain spin systems~\cite{PhysRevLett.93.237204,PhysRevLett.114.177206}, opto-mechanical settings~\cite{aspelmeyerRMP}, superconducting Josephson junctions~\cite{RevModPhys.73.357}, and trapped ions~\cite{ciracPRL}. It exhibits a rich variety of genuinely quantum features, most notably the establishment of quantum correlations between the modes. Thus, the modes share some information that is in principle only accessible from the joint state of both. Our interest lies in understanding the key parameters that determine how accurately we can measure the temperature of the system and designing practical schemes that allow for almost Heisenberg limited temperature estimation.

\section{Preliminaries}
\label{}
\subsection{Classical and quantum estimation strategies}
Information about an unknown parameter, $\mu$, which is imprinted in a quantum system $\rho(\mu)$, can be revealed by measuring any arbitrary observable over the system. 
By repeating such a measurement a large number of times, a dataset of outcomes is collected, upon which one might build up an \textit{estimator} $\hat{\mu}$ in order to estimate the parameter.
Since statistical error---arising from the uncertainty in the outcomes of the measurement---is inescapable, a crucial task in metrology is its identification and optimization.
For any unbiased estimator, i.e. $\big<\hat{\mu}\big>=\mu$, the statistical error is quantified by the (square root of the) variance of the estimator, which according to the Cram\'er-Rao inequality is lower bounded by \cite{frieden2004science,gershenfeld1999nature}
\begin{equation}
\label{bound}
\mbox{Var}(\hat{\mu})\ge\frac{1}{MF(\mu)}.
\end{equation}
Here $M$ denotes the number of measurements employed and $F(\mu)$ the so called Fisher information (FI) associated to the parameter $\mu$. For measurements having a discrete set of outcomes, the FI is given by
\begin{equation}
\label{classical}
F(\mu)=\sum_{j}p_{j}(\partial_\mu\ln p_{j})^2=
\sum_{j}\frac{|\partial_\mu p_j|^2}{p_j},
\end{equation}
where $p_{j}$ represents the probability to get outcome $j$ from the performed measurement. Generalization of Eq.~\eqref{classical} to measurements with continuous outcomes is possible by replacing the summation with an integral, but in the current study we do not deal with such scenarios. As Eq.~\eqref{classical} suggests, the FI can be taken as a measure of \textit{sensitivity} to the parameter:
The larger the FI the more sensitive this measurement is to the unknown parameter, hence the smaller is the statistical error. The dependence of the FI on $p_j$ makes it clear that the quality of the estimation depends on the measurement protocol. However, one may be interested in the ultimate achievable sensitivity, optimized over all possible measurements. This maximum value is called the quantum Fisher Information (QFI)~\cite{lloyd,paris,braunstein_caves,1751-8121-47-42-424006}. The QFI only depends on $\rho(\mu)$, the density matrix of the system and is given by
\begin{equation}
\label{explicit}
{\cal H}(\mu)=\sum_p\frac{[\partial_\mu\rho_p(\mu)]^2}{\rho_p(\mu)}+2\sum_{m\neq n}\sigma_{mn}|\langle\psi_m|\partial_\mu\psi_n\rangle|^2,
\end{equation}
with $\rho_p$ the eigenvalues of the density matrix of the system, $\sigma_{nm}=2\left[\frac{\rho_n(\mu)-\rho_m(\mu)}{\rho_n(\mu)+\rho_m(\mu)}\right]^2$, and $\ket{\psi_i}$ are the eigenstates of the system. Replacing the FI with the QFI in the Eq.~\eqref{classical} gives us the quantum Cram\'er-Rao bound.

In this work we assume our system to have already thermalised to a canonical Gibbs state. The QFI associated to temperature in a thermal state can be simplified by noticing that the eigenstates entering the second term of the RHS of Eq.~\eqref{explicit} do not change with temperature, and therefore this term is identically zero. Thus for a thermal state, the QFI is fully determined by the change of the density matrix eigenvalues with temperature. Equivalently the QFI for thermal states can be determined using~\cite{PhysRevA.78.042105,PhysRevLett.114.220405}
\begin{equation}
\label{varianceH}
{\cal H} = \frac{\Delta\hat H^2}{T^4} = \frac{\big< \hat H^2 \big> - \big< \hat H \big>^2}{T^4}.
\end{equation}
For recent studies on the state-of-the-art regarding thermometry and parameter estimation in thermal states see \cite{ 1367-2630-17-5-055020, 2015arXiv150407787D, PhysRevA.92.052112,PhysRevA.88.063609, ParisJPA, PhysRevA.78.042105,PhysRevA.78.042106,PhysRevA.76.062318,Mehboudi_in_preparation,2017arXiv170108531D,2017arXiv170303719H} and references therein.

\subsection{The model: Coupled non-linear harmonic oscillators}
We consider two interacting bosonic modes as our system, with the Hamiltonian given by
\begin{eqnarray}
\label{hamiltonian}
\hat{H}_0&=& \hbar\omega_1 \left(\hat a^\dagger_1 \hat a_1+\frac{1}{2} \right) + \hbar\omega_2 \left(\hat a^\dagger_2 \hat a_2+\frac{1}{2} \right) - \hbar{\cal J} \left(\hat a^\dagger_1 \hat a_2 + \hat a_1 \hat a^\dagger_2\right) \nonumber \\ 
&+& \frac{\hbar{\cal U}}{2} \left(\hat a^{\dagger2}_1\hat a^2_1+\hat a^{\dagger2}_2\hat a^2_2 \right),
\end{eqnarray}
where the first two terms correspond to the free evolution of each mode, the third term is the inter-mode interaction term, and the final term characterises the non-linearity. In what follows we rescale the Hamiltonian, $\hat{H}_0$ with respect to $\hbar\omega_1$. Therefore we consider the dimensionless Hamiltonian $\hat{H}_0/(\hbar\omega_1)=\hat{H}$ with $\hbar{\cal J}/(\hbar\omega_1)\to J$, $\hbar{\cal U}/(\hbar\omega_1) = U$, and $\hbar\omega_2/(\hbar\omega_1) = (1+\Delta)$. Throughout our study we assume the system to be in a thermal state due to interaction with its environment. Thus, our system will be described by the Gibbs state of the form
\begin{equation}
\label{gibbs}
\rho(\beta)=\frac{e^{-\beta \hat{H}}}{\cal Z},
\end{equation}
with ${\cal Z}$ the associated partition function and for compactness of notation here, and throughout, we have used $\beta=\frac{1}{k_B T}$. The central aims of this work concern finding the ultimate bounds on precision of estimating $T$ (or equivalently $\beta$), the difficulties to achieve such a precision, and analysing alternative scenarios that despite returning sub-optimal precisions, are more experimentally viable. The interaction with the thermal bath leading to Eq.~\eqref{gibbs}, can be with or without particle exchange. Sec.~\ref{DWs} is dedicated to interactions which commute with the total number of bosons, hence leaving it a conserved quantity. This case resembles a double-well potential loaded with a Bose-Einstein condensate. In Sec.~\ref{harmosc} we consider thermal states in which the total number of bosons in the two modes is not fixed. In this case the model can be used to study, e.g. coupled opto-mechanical systems and trapped ions.

\section{Two mode Bose-Hubbard model}
\label{DWs}
We begin by fixing the total number of bosons $N=a^\dagger_1 a_1 + a^\dagger_2 a_2 $ as a constant, thus Eq.~\eqref{hamiltonian} is the familiar two-site Bose-Hubbard model and the system is effectively a double-well potential loaded with a Bose-Einstein condensate. In this context $J$ plays the role of the tunnelling, while $U$ encompasses the self-interaction of the atoms within each well, and $\Delta$ determines if the two wells are resonant.

To begin we set $\Delta=0$ such that the wells are on resonance. For small self-interactions, $NU\ll J$, the model can be solved by mapping Eq.~\eqref{hamiltonian} to a harmonic oscillator through the Holstein-Primakoff (HP) approximation. We do this by introducing the Schwinger operators
\begin{eqnarray}
\hat S_x = \frac{1}{2}\left( \hat a_1^\dagger \hat a_1 - \hat a_2^\dagger \hat a_2 \right),\\
\hat S_z = \frac{1}{2}\left( \hat a_1^\dagger \hat a_2 + \hat a_2^\dagger \hat a_1 \right),
\end{eqnarray}
and, for simplicity neglecting the free evolution terms, we can re-write Eq.~\eqref{hamiltonian} to be 
\begin{equation}
\label{LMG}
\hat H = -2 J \hat S_z + U\hat S_x^2, 
\end{equation}
(note we have excluded a constant term $U \left( \frac{N^2}{4} - \frac{N}{2} \right)$ that is also immaterial to our analysis). Through the HP transformation~\cite{PhysRevLett.93.237204,PhysRevLett.114.177206} 
\begin{equation*}
\hat S_x = \frac{\sqrt{N}}{2}\left( \hat a+ \hat a^\dagger \right),~~~{\rm and}~~~~
\hat S_z = \frac{N}{2} - \hat a^\dagger \hat a,
\end{equation*}
and with the suitable Bogoliubov transformation
\begin{equation*}
\hat a=\sinh\left(\frac{\alpha}{2}\right) \hat b^\dagger + \cosh\left(\frac{\alpha}{2}\right) \hat b,~~{\rm and}~~
\hat a^\dagger=\sinh\left(\frac{\alpha}{2}\right) \hat b + \cosh\left(\frac{\alpha}{2}\right) \hat b^\dagger.
\end{equation*}
with $\tanh\alpha=-\frac{UN}{4J+UN}$, we can map the original Hamiltonian into
\begin{equation}
\label{DWmapped}
\hat H = \omega \left (\hat b^\dagger \hat b + \frac{1}{2} \right ) -J(N+1)~~~{\rm with}~~\omega=\sqrt{2 J (2 J + N U)}.
\end{equation}
With this mapping, determining the QFI for temperature reduces to evaluating the variance of this Hamiltonian and plugging it in Eq.~\eqref{varianceH}~\cite{PhysRevLett.114.220405}. We find it takes the concise analytic form
\begin{equation}
\label{DWQFI}
\mathcal{H} = \frac{\beta^4 J}{2}\left( 2J+N U \right) {\rm csch}^2 \left( \beta \sqrt{J\left( J + \frac{N U}{2} \right)} \right).
\end{equation}
We remark that this expression is valid when $\big<\hat b^\dagger \hat b\big> \ll N$.

\begin{figure}[t]
\begin{center}
{\bf (a)}\hskip0.5\columnwidth{\bf (b)}\\
\includegraphics[width=0.5\columnwidth]{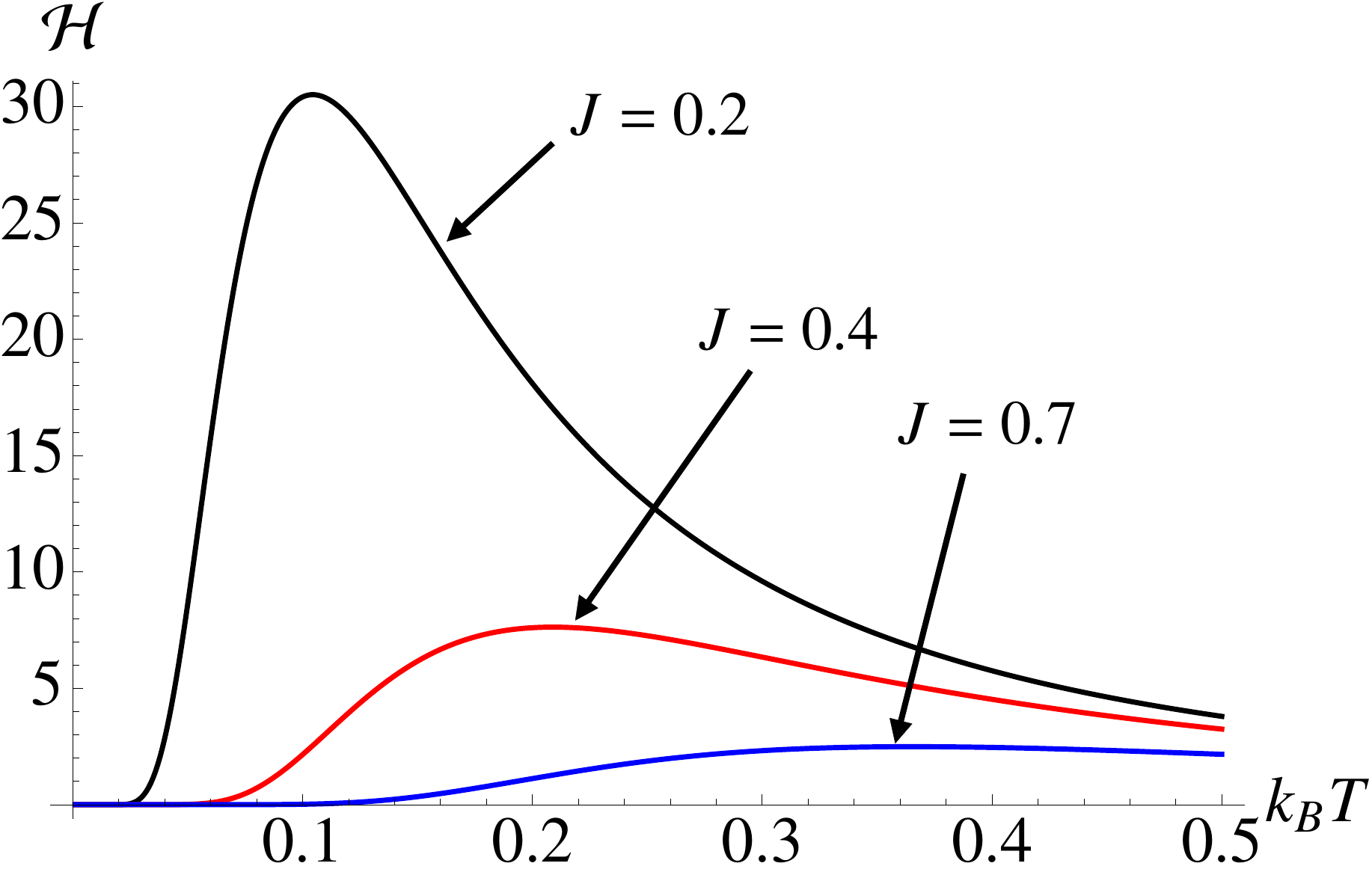}~\includegraphics[width=0.5\columnwidth]{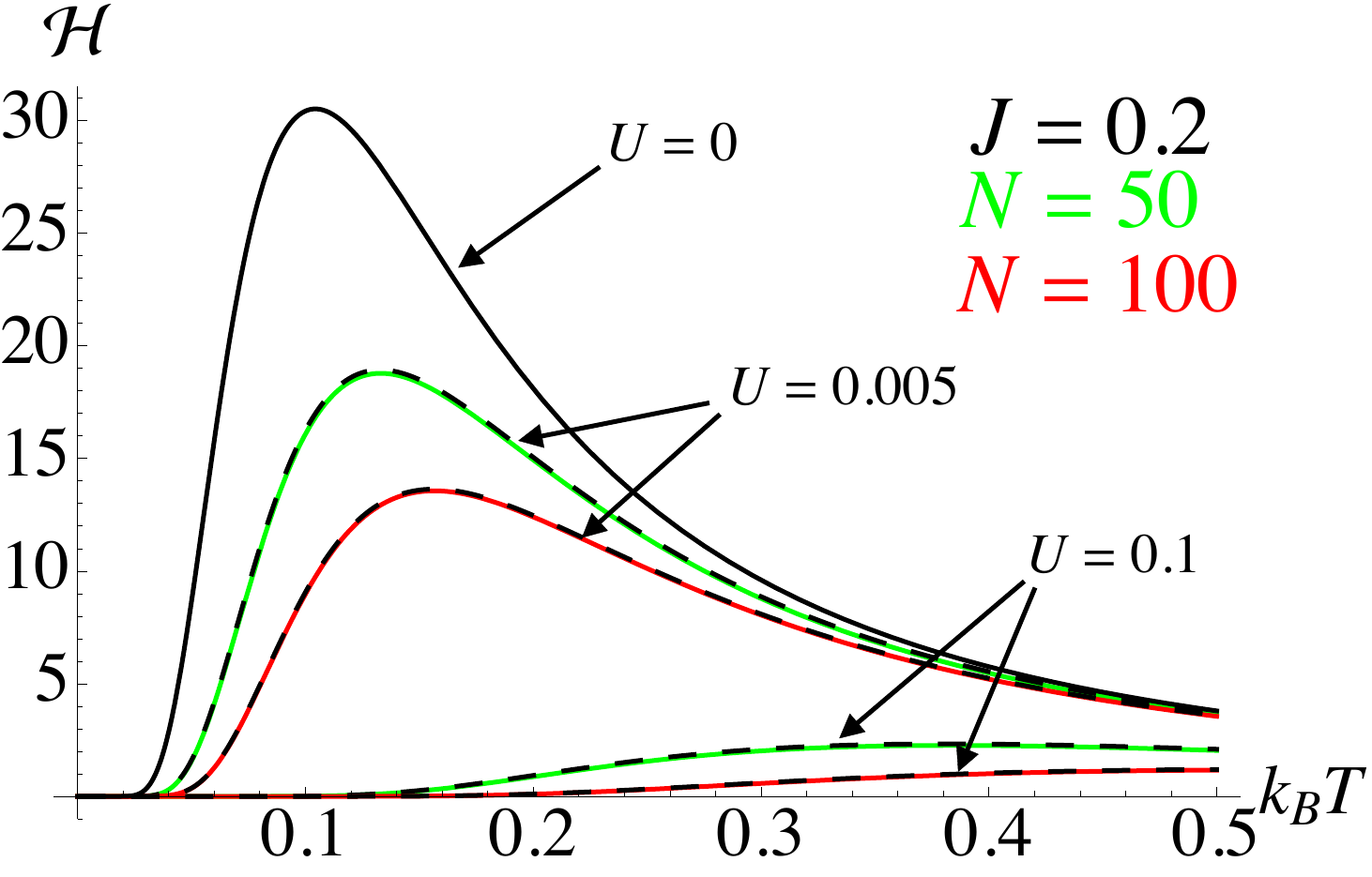}
\end{center}
\caption{QFI for a double-well potential. {\bf (a)} Taking $U=0$, we examine the behaviour for difference values of the tunnelling strength. {\bf (b)} Examining the behaviour for different $U$ and atom number, taking $N=$50 (green) or 100 (red) and fixing $J=0.2$. The top-most black curve is for $U=0$ and we consider $U=0.005$ and $U=0.1$. The dashed black curves are the numerically evaluated QFI while the solid coloured lines are for the analytic expression from the HP approximation, Eq.~\eqref{DWQFI}.}
\label{fig1N}
\end{figure}

In Fig.~\ref{fig1N} {\bf (a)} we plot the behaviour of $\mathcal{H}$ for several values of $J$ in the tunnelling dominated regime, i.e. $U\!\!=\!\!0$. We see that an increase in the tunnelling {\it decreases} the QFI. Furthermore, in this regime the estimation of temperature is independent of the value of $N$, i.e. no advantage or disadvantage appears by using larger or smaller numbers of atoms, and this is due to the fact that none of the atoms are interacting with one-another. Turning our attention to non-zero self interactions, in panel {\bf (b)} we see even small values of $U$ can have quite drastic effects and now the system is clearly dependent on the number of atoms $N$. The ability to accurately estimate the temperature decreases both with increasing system size and self-interaction strength. 

We can understand this behaviour by examining the distribution of energy levels. From $\omega$ given in Eq.~\eqref{DWmapped}, it is clear that in the tunnelling dominated regime, the effective frequency of the mapped harmonic oscillator depends only on $J$, and hence the invariance to system size, $N$, is readily understood. Furthermore, as we increase $J$ the spacing between adjacent energy levels increases. Therefore, for small but non-zero $J$ the energy spacing between the levels is also small, and so it requires less thermal energy to occupy higher eigenstates. As the QFI for thermal states is entirely determined by the rate of change of these occupations it means that the smaller the tunnelling the more accurately we can estimate the temperature. Notice that regardless of the value of $J$ in {\bf (a)} we see there is a range of temperatures where $\mathcal{H}$ is zero and no estimation of the temperature is possible. This is because for all non-zero $J$ there is a finite gap between the ground and first excited states. If the temperature of the state is insufficient to excite any occupation to the first excited state the system remains in its ground state, despite being at a finite temperature. As there is only single occupation, no variance is present and the QFI is correspondingly zero. 

In panel {\bf (b)} we take finite values for the self-interaction term. While many of the qualitative features discussed for the $U\!\!=\!0$ case still hold, we see the overall effect that $U$ has is to degrade the sensitivity to temperature variation of the system. This behaviour is again readily explained by examining the effect finite $U$ has on the energy level spacing as dictated by $\omega$ in Eq.~\eqref{DWmapped}. When $N$ is suitably large, such that the mapping is meaningful, even a small value of $U$ can significantly increase the gap between energy levels, which in turn leads to a decrease in the QFI.

We remark that Eq.~\eqref{DWQFI} is exact for $U\!\!=\!0$, however otherwise it is an approximation accurate up to the validity of the HP mapping. With this in mind, in Fig~\ref{fig1N} {\bf (b)} the dashed lines correspond to the numerically evaluated QFI found by computing the thermal state of Eq.~\eqref{LMG} and directly diagonalising, thus confirming that we are in a regime where the mapping holds. The overall effect of $\Delta\neq0$ is to reduce the QFI, the reasons for which will be addressed in the following section as the effect of detuning on the spectrum when we do or do not allow for particle exchange with the bath is the same. 

\section{Coupled harmonic oscillators in a common bath}
\label{harmosc}
We next relax the assumption that the total number of bosons is fixed and instead allow for the system to exchange particles with the thermal bath. Thus we shift paradigm and consider coupled harmonic oscillators. This system encompasses many physically relevant situations such as ultra-cold atoms, trapped ions, and some important light-matter systems such as opto-mechanics~\cite{aspelmeyerRMP}. The term $J$ in this setting corresponds to the oscillator coupling strength, while $\Delta$ is a detuning, and $U$ encompasses a non-linear term.

\subsection{Heisenberg limited temperature estimation}
When the non-linear term $U=0$, due to the Gaussian nature of the model it can be solved analytically. In this case, our Hamiltonian can be rewritten in the operator basis $\{\hat x_1,\hat p_1,\hat x_2,\hat p_2\}$ where
\begin{equation}
\hat x_j=\frac{1}{\sqrt{2}}\left( \hat a_j + \hat a^\dagger_j  \right),~~~{\rm and}~~~\hat p^\dagger_j=\frac{i}{\sqrt{2}}\left( \hat a^\dagger_j - \hat a_j  \right).
\end{equation}
We can straightforwardly diagonalise $\hat{H}$, see for example Ref.~\cite{campbellSciRep}, and express it in diagonal form in terms of the two normal modes, with frequencies given by
\begin{equation}
\omega_\pm=\frac{1}{2} \left(2+\Delta \pm \sqrt{\Delta^2 +4 J^2}\right).
\end{equation}
All information on the state of the system contained in Eq.~\eqref{gibbs} is equivalently contained in the associated covariance matrix $\sigma$ of entries $\sigma_{ij}=\frac{1}{2}\langle\{\hat P_i,\hat P_j\}\rangle-\langle \hat P_i\rangle\langle \hat P_j\rangle$, where $\hat P_i$'s are the elements of the vector of quadrature operators $\hat P^\top=(\hat x_1~\hat p_1~\hat x_2~\hat p_2)$ and the expectation value of such a vector (calculated over the state of the system). We find our covariance matrix is
\begin{equation}
{\bm \sigma}_\Delta(\beta)=\frac{1}{\mathcal{N}}\left(
\begin{array}{cccc}
 K^+ & 0 & 2J \sinh ( C ) & 0 \\
 0 & K^+ & 0 & 2J \sinh ( C ) \\
2J \sinh ( C ) & 0 & K^- & 0 \\
 0 &2J \sinh ( C ) & 0 & K^- \\
\end{array}
\right),
\end{equation}
with $\mathcal{N}=A \left(\cosh ( B )-\cosh ( C ) \right)$, $K^\pm=A \sinh ( B ) \pm \Delta \sinh ( C )$, $A=\sqrt{\Delta^2+4J^2}$, $B=\frac{1}{2}\beta(2+\Delta)$, and $C=\frac{1}{2} \beta\sqrt{\Delta^2+4J^2}$. Despite not being available analytically, when explicitly considering the case of $U\neq0$ we can determine the thermal state Eq.~\eqref{gibbs}, and thus the QFI, numerically. 

For $U=0$, and in the normal mode representation, the system is described by a product state of two uncoupled oscillators with frequencies $\omega_{\pm}$, and its density matrix,
\begin{eqnarray}
	\rho(\beta)=\frac{e^{-\beta~H^{+}}}{\mathrm{Tr}[e^{-\beta~H^{+}}]}\otimes\frac{e^{-\beta~H^{-}}}{\mathrm{Tr}[e^{-\beta~H^{-}}]}.
\end{eqnarray}
Here we define $H^{\pm}$ to be the free Hamiltonian of a harmonic oscillator with the normal mode frequency $\omega_{\pm}$.
On this account, and by using the additivity of the quantum Fisher information for product states \cite{1751-8121-47-42-424006}, we find that
\begin{equation}
	{\cal H}=\frac{\beta^4}{4}\left(\omega_{+}^2\mathrm{csch}^2\frac{\beta\omega_+}{2}+\omega_{-}^2\mathrm{csch}^2\frac{\beta\omega_{-}}{2}\right),
\end{equation}
where we use the fact that for a single harmonic oscillator with frequency $\Omega$, the QFI is given by ${\cal H}^{ho}=(\beta^4\Omega^2/4)\mathrm{csch}^2(\beta\Omega/2)$ \cite{PhysRevLett.114.220405}. Before proceeding further let us remark that for a fixed temperature ${\cal H}^{ho}$ decreases monotonically by increasing $\Omega$. Therefore, a harmonic oscillator with a smaller frequency is more sensitive to temperature.

\begin{figure}[t]
\begin{center}
{\bf (a)}\hskip0.45\columnwidth{\bf (b)}\\
\includegraphics[width=0.45\columnwidth]{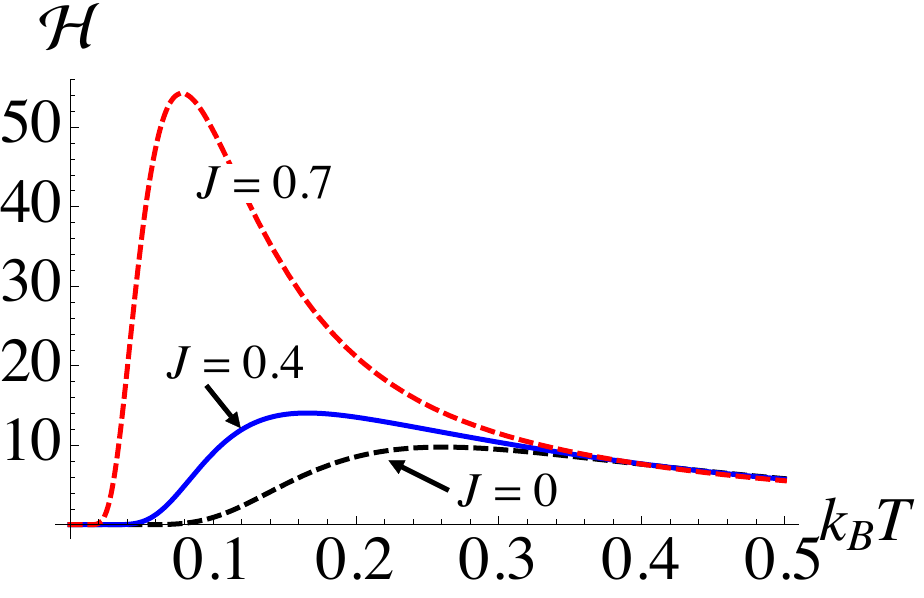}~\includegraphics[width=0.45\columnwidth]{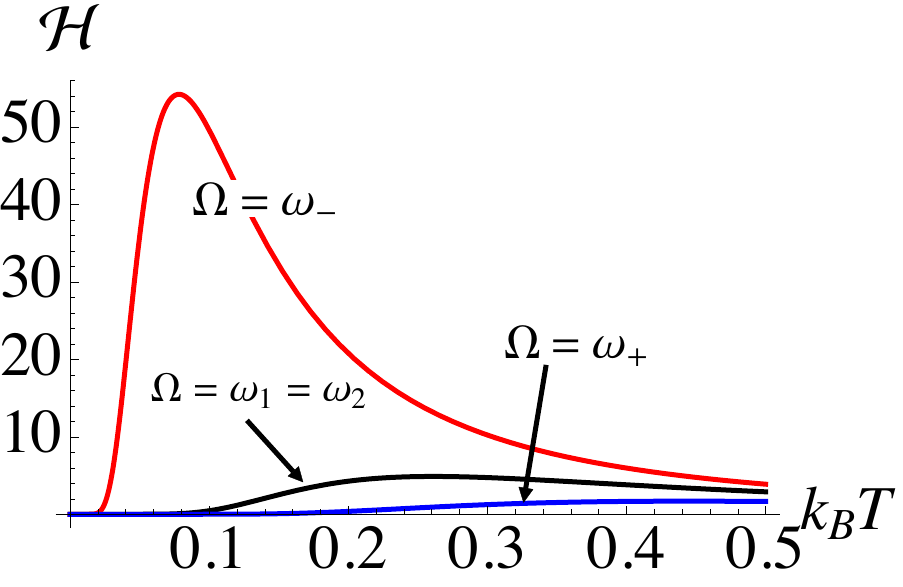}\\
{\bf (c)}\hskip0.45\columnwidth{\bf (d)}\\
\includegraphics[width=0.45\columnwidth]{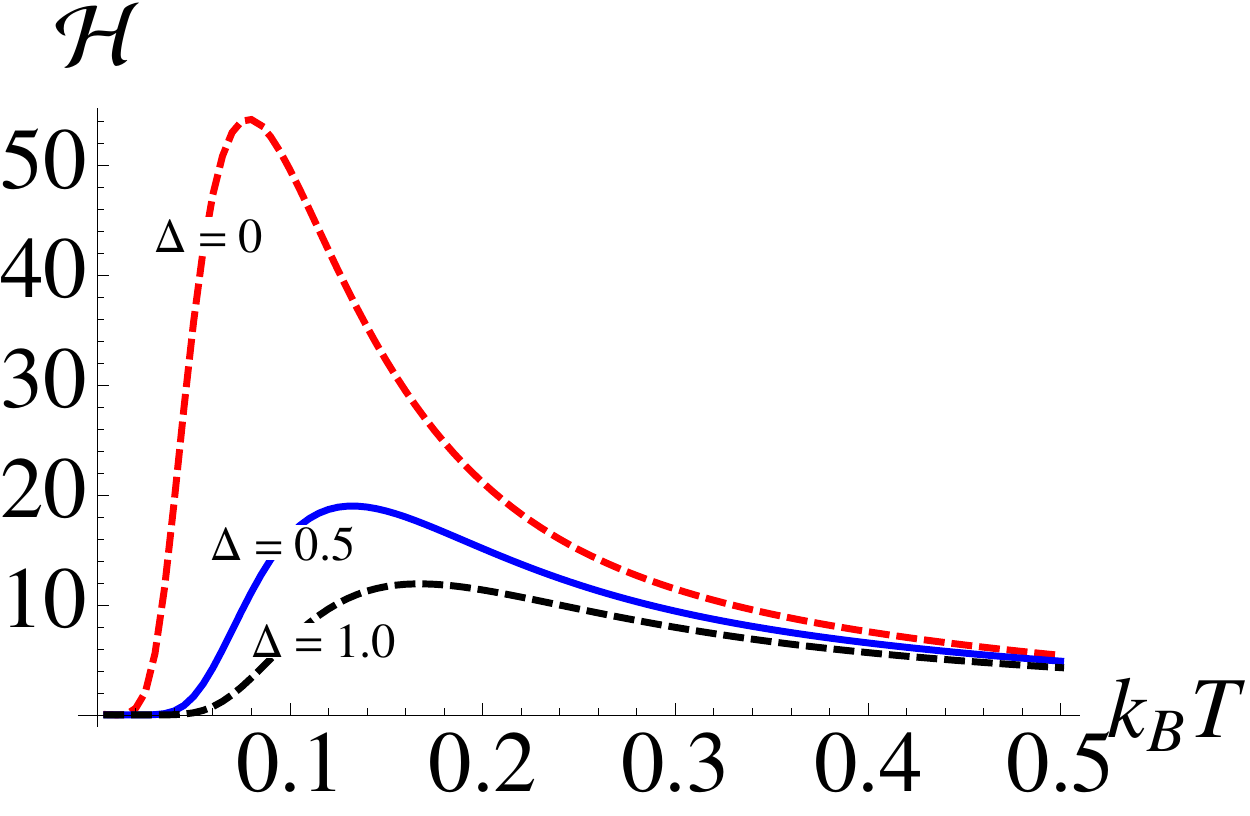}~\includegraphics[width=0.45\columnwidth]{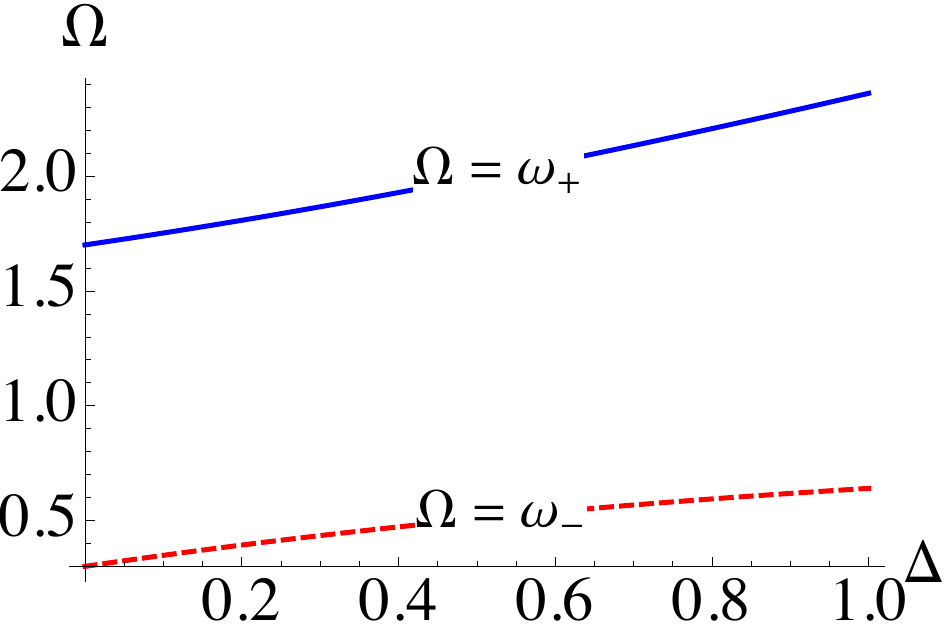}\\
{\bf (e)}\hskip0.45\columnwidth{\bf (f)}\\
\includegraphics[width=0.45\columnwidth]{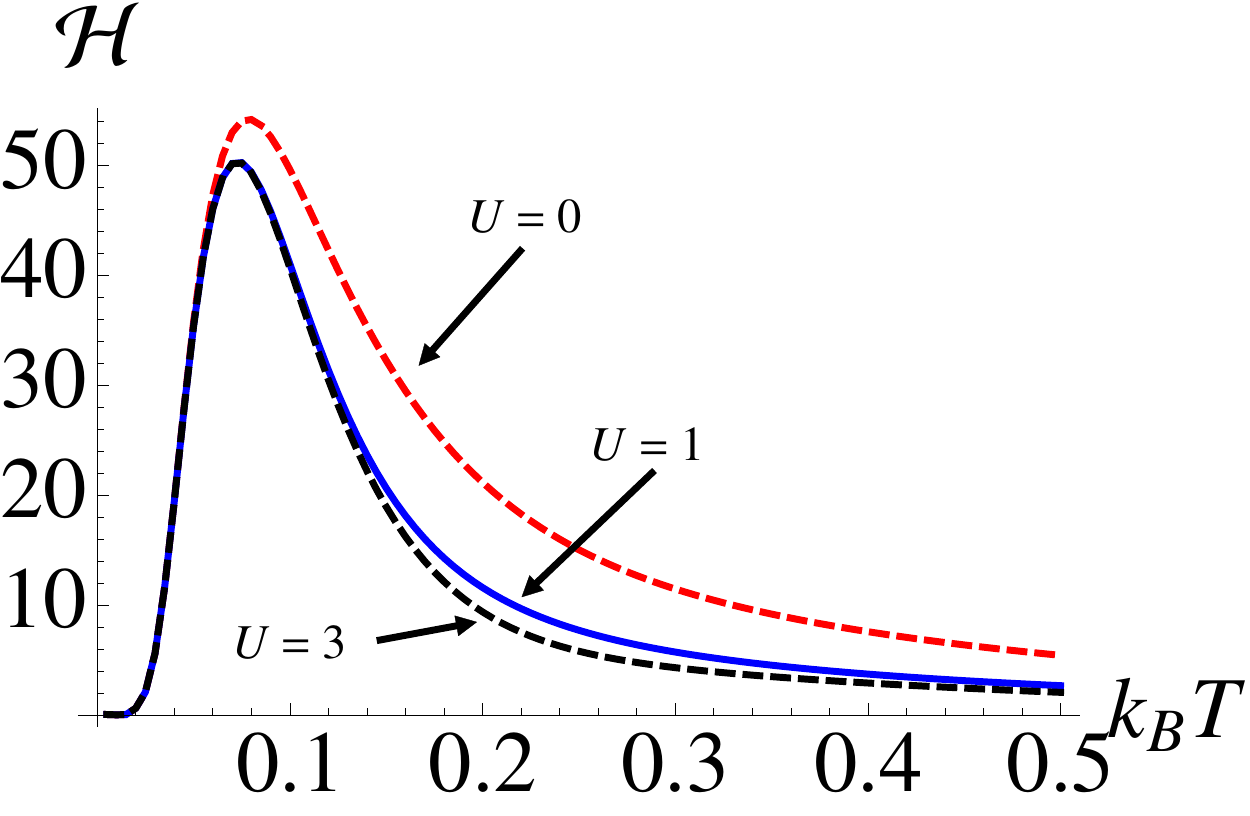}~\includegraphics[width=0.45\columnwidth]{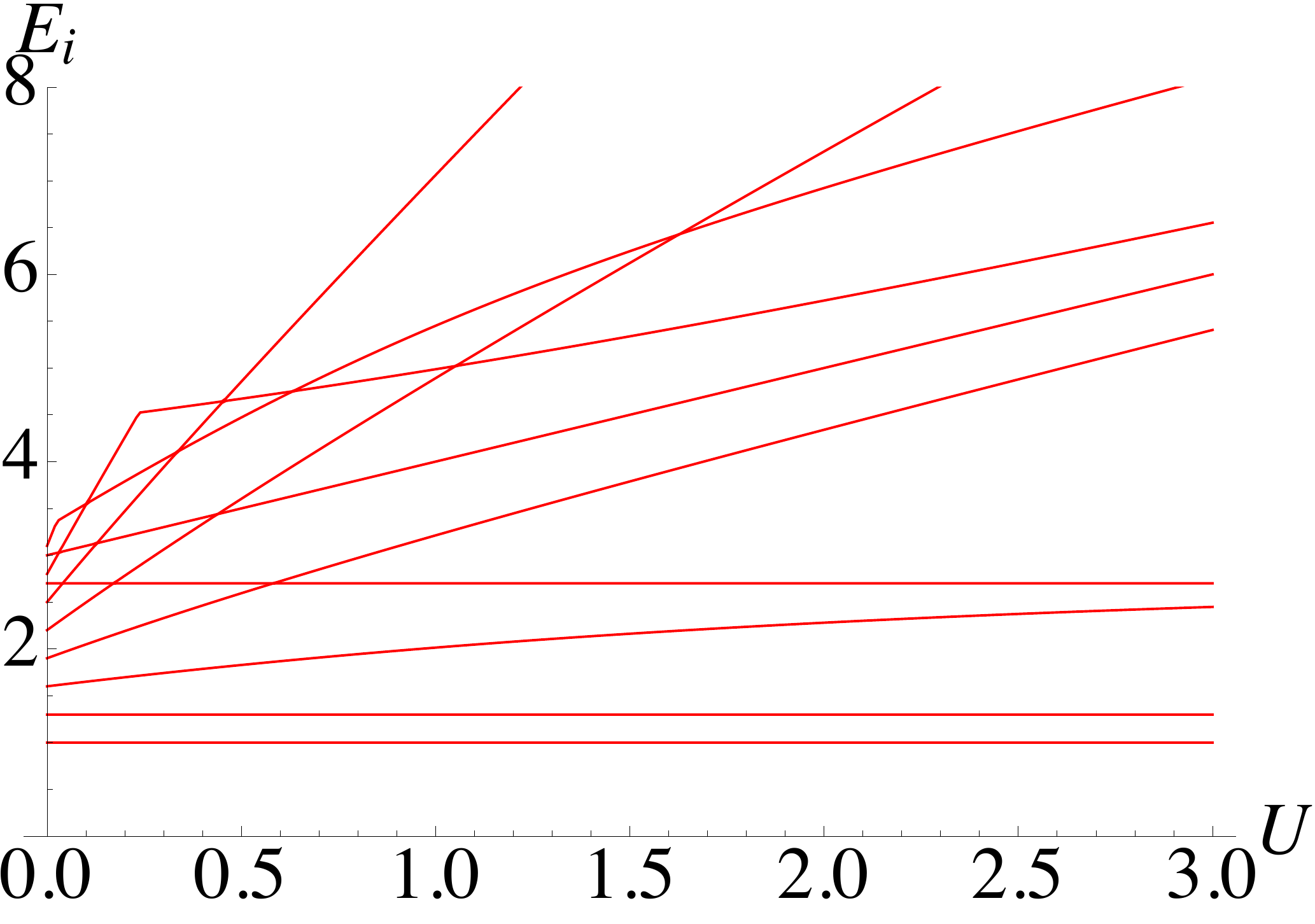}
\end{center}
\caption{{\bf (a)} The QFI of the \textit{coupled} oscillators against $k_BT$, fixing $U=\Delta=0$ and looking at different values of the coupling strength $J=\{0, 0.4,0.7\}$.
{\bf (b)} The QFI of a \textit{single} oscillator for different frequencies $\Omega=\{\omega_1=\omega_2,\omega_{-},\omega_+\}$.
{\bf (c)} The QFI of the \textit{coupled} oscillators against $k_BT$, fixing $J=0.7$, $U=0$ and examining different values of detuning $\Delta$.
{\bf (d)} The normal mode frequencies against $\Delta$, for $J=0.7$ and $U=0$.
{\bf (e)} The QFI of the \textit{coupled} oscillators against $k_BT$, fixing $J=0.7$, $\Delta=0$ and examining different values of non-linearity $U$.
{\bf (f)} Energy spectra of the lowest 10 levels of the coupled oscillators against $U$ with $J=0.7$ and $\Delta=0$.}
\label{fig2N}
\end{figure}

In Fig.~\ref{fig2N} {\bf (a)} we depict the QFI of the coupled harmonic oscillators versus temperature. We set $\Delta=U=0$ and look at different values of $J$. Interestingly, contrary to the behaviour for the loaded double-well, we observe that by increasing the interaction strength the thermometry precision enhances. In particular, this enhancement is more significant at lower temperatures, a regime where thermometry precision is known to be a challenging task \cite{correa2016low}. This behaviour can be understood with the help of the normal mode frequencies. By increasing the interaction strength $J$, the normal frequency $\omega_{-}$ decreases, hence making it a much more precise thermometer. On the other hand, the other normal mode frequency, $\omega_{+}$, increases with $J$, hence, it effectively becomes a less sensitive thermometer. Nevertheless, the improvement attained from decreasing $\omega_{-}$ is so large that it not only compensates for $\omega_{+}$, it makes the total system a notably more sensitive thermometer as well, as can be clearly seen from Fig.~\ref{fig2N} {\bf (b)}.

Next, we explore the impact of detuning on thermometry precision. To this aim, in Fig.~\ref{fig2N} {\bf (c)}, by fixing $J=0.7$ and $U=0$, we plot the QFI against temperature, for different values of detuning. We see that as the oscillators are taken more off-resonance the QFI rapidly decreases. 
Again this can be explained with the help of normal mode frequencies. It is easy to verify that both frequencies are monotonically increasing functions of $\Delta$, for any value of $J$. That being the case, and due to the fact that for a fixed temperature ${\cal H}^{ho}$ monotonically decreases with frequency, we deduce that increasing the detuning leads to a drop in thermometry precision, cf. Fig.~\ref{fig2N} {\bf (d)}.

Finally, in Fig.~\ref{fig2N} {\bf (e)} we examine the effect of the self interaction term by numerically evaluating Eq.~\eqref{explicit} and Eq.~\eqref{gibbs}. Fixing $J=0.7$ and $\Delta=0$, we see that as we increase $U$, the QFI (in general) decreases. However, the quantitative decrease in this case is not very significant, which is again in stark contrast to the behaviour for the loaded double-well. 
Insight into this is found by examining the energy spectrum [cf. Fig.~\ref{fig2N} {\bf (f)}]. The two lowest energy levels are insensitive to the change in $U$, only from the second excited state onwards are the energy eigenvalues affected. Therefore, as we see in Fig.~\ref{fig2N} {\bf (c)}, for an initial window in $k_B T$ the QFI remains the same regardless of the magnitude of $U$. As the temperature is increased, and hence the high-energy states start to be populated, the slightly larger gap between energy levels due to the non-zero values of $U$ is reflected in a decrease in the maximal value of the QFI. However, as $U$ has a comparatively weak effect on the second and third excited states compared to the higher energy levels, the overall impact that it has is quite small in the low temperature range. 

In summary, we have shown that the magnitude of the interaction is the dominant parameter in maximising the QFI. At low temperatures, the non-linearity $U$ plays little or no role in the overall ability to estimate the temperature, likely in light of the fact that at low temperature the anharmonicity of the corresponding oscillators is less evident (as only low-energy states will be involved in the decomposition of the state of the system). The non-linear effects become more pronounced as $T$ increases. Furthermore, ensuring that the two oscillators are on-resonance is shown to be a vital feature. In what follows we will therefore assume $U=\Delta=0$.

\section{Local schemes for temperature estimation}
While the QFI provides us with the upper bound for estimating the temperature, it does not give any indication of how an implementable measurement approach would perform. In order to assess this, we must choose an {\it experimentally viable} measurement strategy and determine the post-measurement classical FI, Eq.~\eqref{classical}. This will allow us to quantitatively examine how close the FI gets to the upper bound given by the QFI. 

Arguably the least experimentally demanding approach will be to restrict to measurements on only one of the two oscillators and we consider two typical measurement strategies: (i) homodyne and (ii) local energy measurements. A remark, since we are assuming both oscillators are on resonance it is immaterial which is chosen to be measured. Homodyne measurements are achieved by projecting the state onto the eigenstates of the quadrature operator $\hat q_1(\theta)=(\sqrt 2)^{-1/2} \left(\hat a_1 e^{i\theta}+\hat a^\dag_1 e^{-i\theta} \right)$, with $\theta\in[0,2\pi)$. Taking $\theta=0$ corresponds to measurements of the position quadrature $\hat x_1$. However, such homodyne measurements are found to be very ineffective [plots not shown], achieving a FI of less than half the QFI. 

\begin{figure}[t]
\begin{center}
{\bf (a)}\hskip0.5\columnwidth{\bf (b)}\\
\includegraphics[width=0.45\columnwidth]{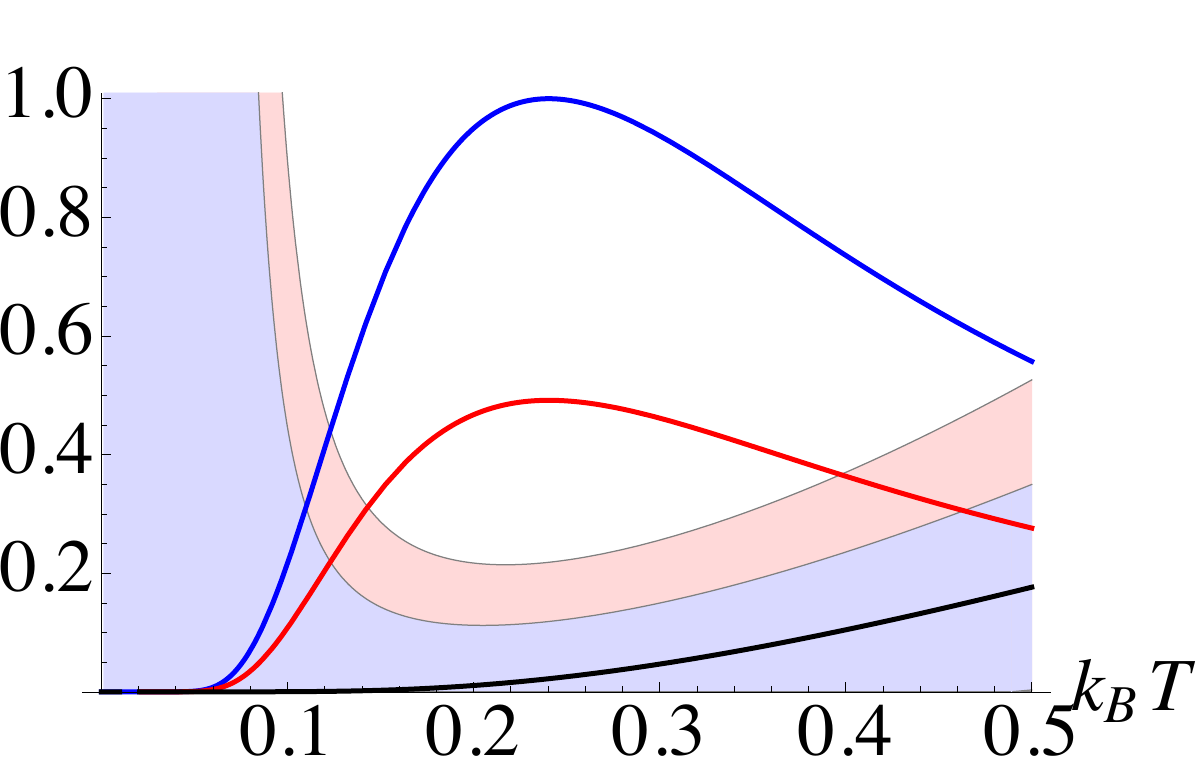}~\includegraphics[width=0.45\columnwidth]{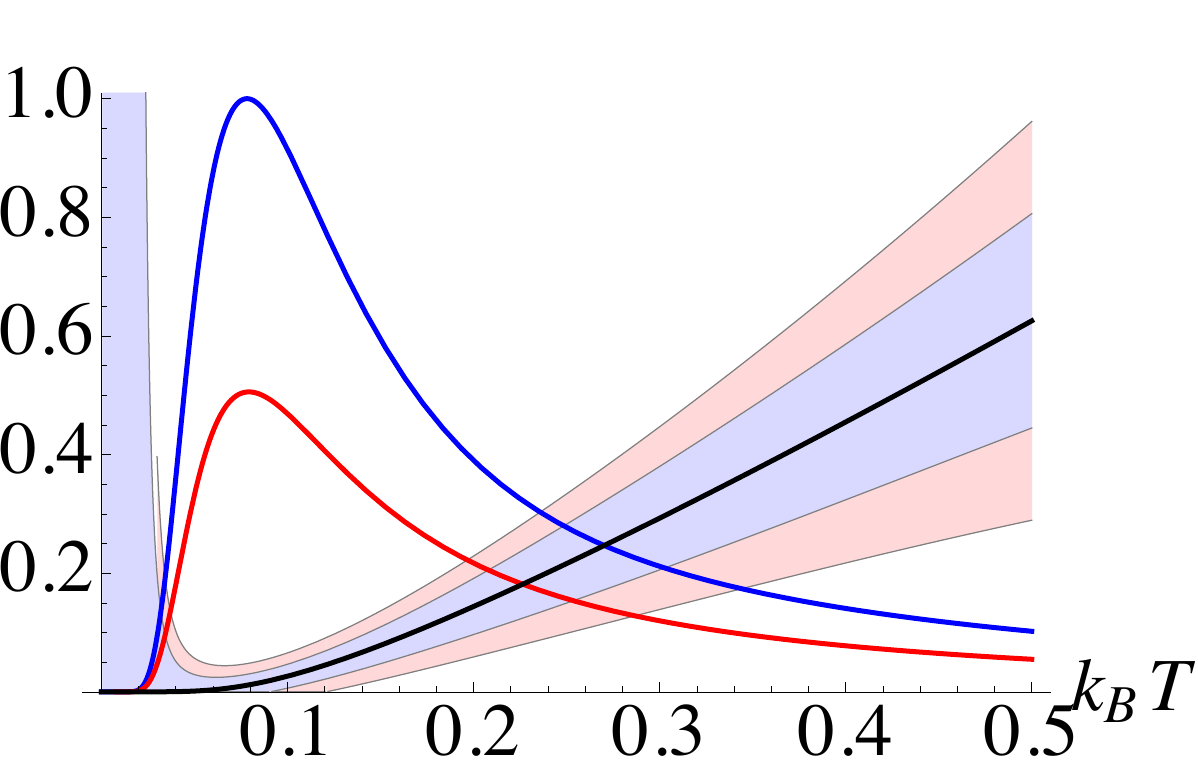}\\
{\bf (c)}\hskip0.5\columnwidth{\bf (d)}\\
\includegraphics[width=0.45\columnwidth]{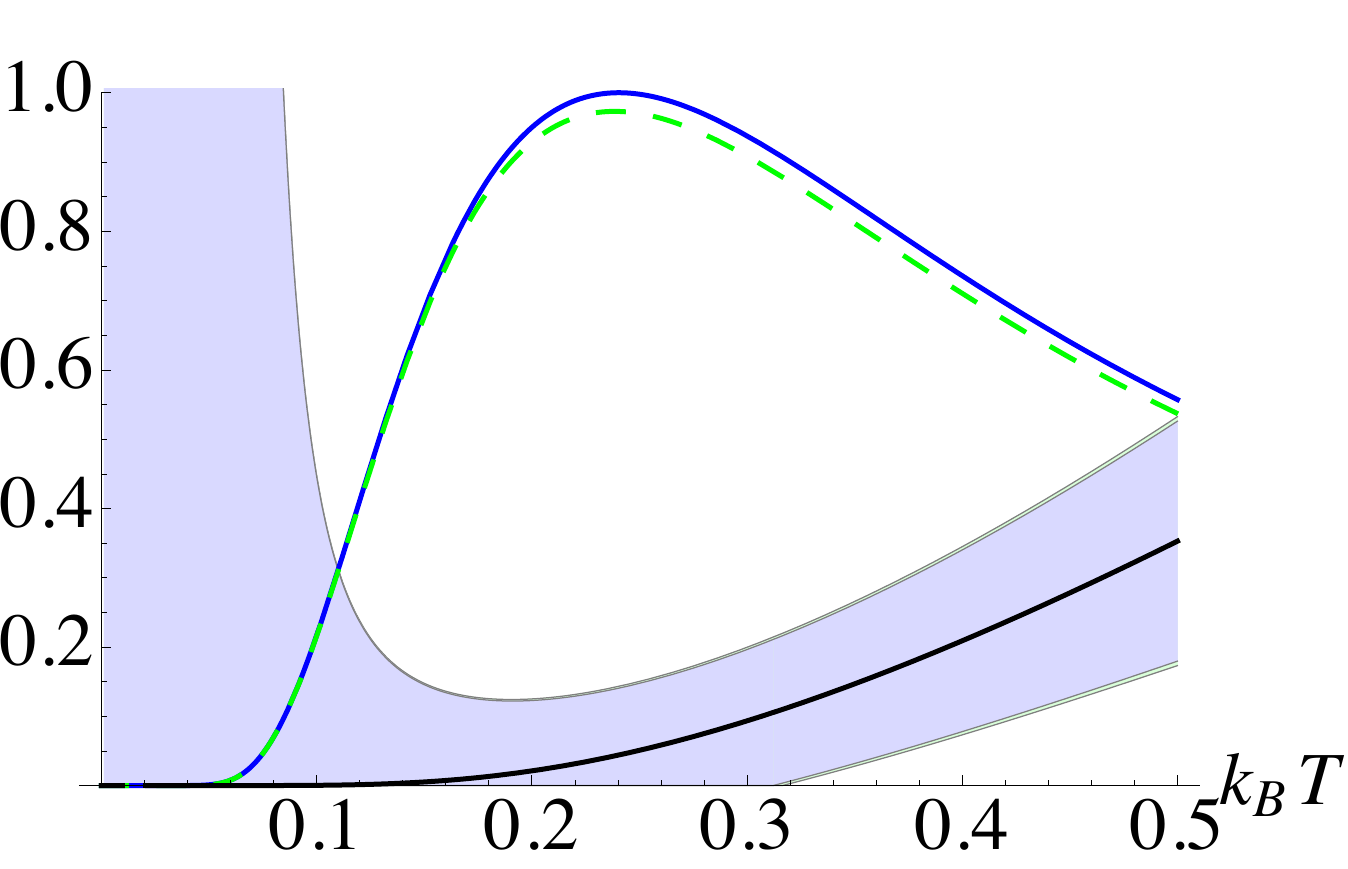}~\includegraphics[width=0.45\columnwidth]{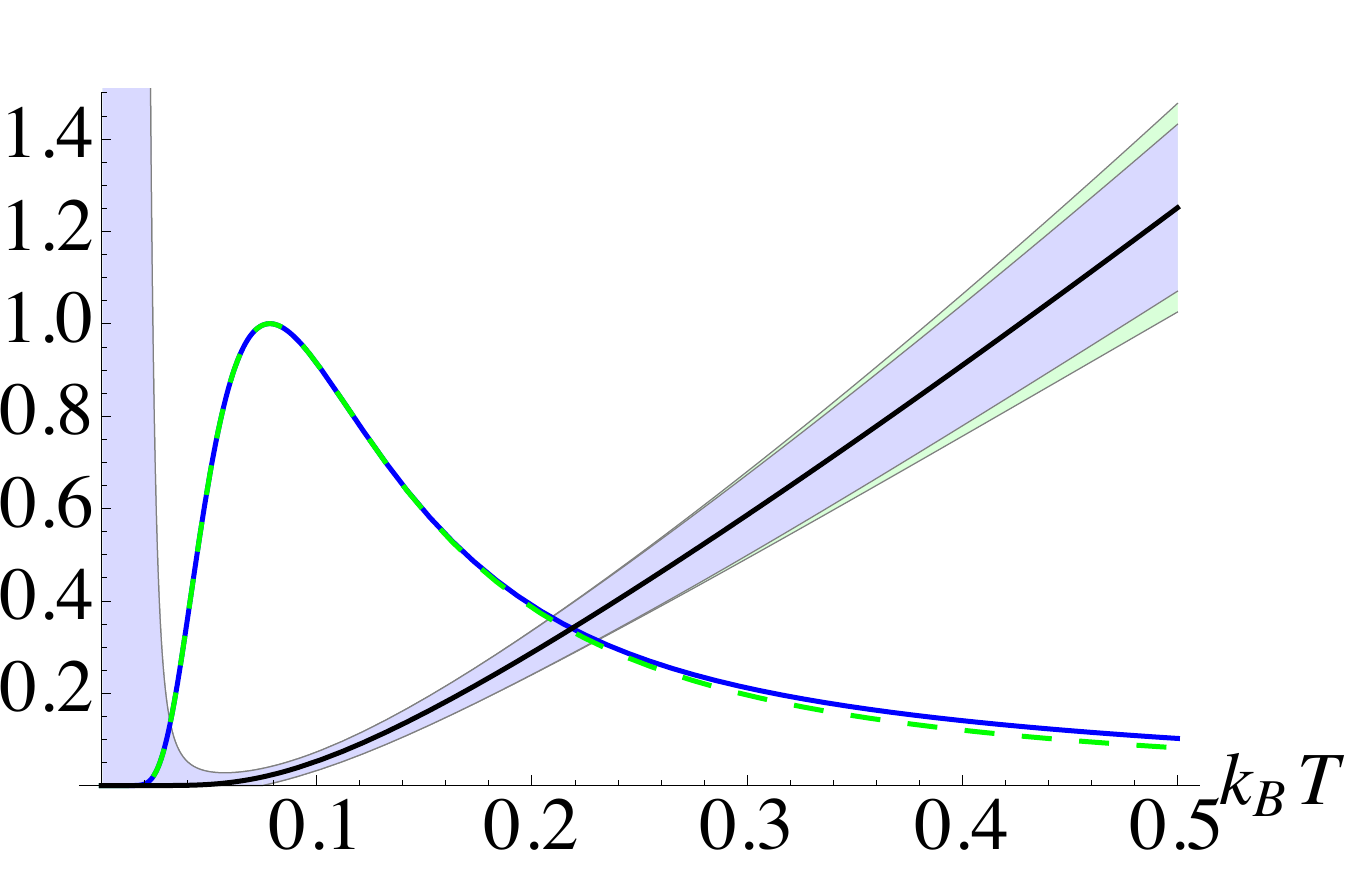}
\end{center}
\caption{{\bf (a)} and {\bf (b)} Rescaled QFI (blue) ($\mathcal{H}/\mathcal{H}_{\rm max}$) and rescaled FI (red) ($F_{n_1}/\mathcal{H}_{\rm max}$) for the energy measurement on one oscillator, the black curve corresponds to $\big<\hat{a}_1^\dagger \hat{a}_1\big>$. The inner-blue shaded region shows the quantum Cram\'er-Rao bound and the outer-red shaded region shows the Cram\'er-Rao bound corresponding to this measurement. Each panel corresponds to an increasingly large value of the tunnelling strength, $J=0.2$ and 0.7 respectively. {\bf (c)} and {\bf (d)} As for the previous panels, however here the green-dashed curve corresponds to the $F_{n_1+n_2}/\mathcal{H}_{\rm max}$ for the `global' measurement of the expectation value $\big<\hat{a}_1^\dagger \hat{a}_1+\hat{a}_2^\dagger \hat{a}_2\big>$ (black curve), and the outer-green shaded area the associated Cram\'er-Rao bound.}
\label{fig3N}
\end{figure}

Energy measurements are simply achieved by projecting onto the eigenstates of the Hamiltonian for the free evolution of one oscillator, i.e.
\begin{equation}
\hat{H}_1=\left(\hat a^\dagger_1 \hat a_1+\frac{1}{2} \right).
\end{equation}
As we know, there is a one-to-one relation between the population measurement $\big< \hat{a}_1^\dagger \hat{a}_1\big>$ and $\beta$. Using this strategy we can find the FI is
\begin{equation}
F_{n_1} = \frac{\beta ^4 \left\{ J \sinh (\beta ) \sinh (\beta  J)-\cosh (\beta ) \cosh (\beta  J)+1 \right\}^2}{\left\{ \cosh (\beta )-\cosh (\beta  J)\right\}^2 \left\{\sinh ^2(\beta )-\left[ \cosh (\beta )-\cosh (\beta  J)\right]^2\right\}}.
\end{equation}
In Fig.~\ref{fig3N} {\bf (a)} and {\bf (b)} we show how this measurement approach performs. A remark, if $J=0$ we find that $F_{n_1} = \frac{1}{2} {\cal H}$---because the two oscillators are uncoupled and additivity of the QFI holds---while for $J\neq0$ we see $F_{n_1} > \frac{1}{2} {\cal H}$, showing that even with simple local measurements the interaction term allows for more accurate determination of the temperature.

Relaxing the constraint on measuring only one oscillator, we find a `near optimal' performance can be achieved with a global (albeit, somewhat trivially global) measurement: $\big< \hat{a}_1^\dagger \hat{a}_1 + \hat{a}_2^\dagger \hat{a}_2 \big>$, and we can evaluate the FI analytically from the covariance matrix elements finding
\begin{equation}
F_{n_1+n_2} = \frac{\beta^4 \left\{ 1 - \cosh (\beta) \cosh (\beta J) + J \sinh (\beta) \sinh (\beta J) \right\}^2}{\{\cosh(\beta)-\cosh(\beta J)\}^2\{\cosh(\beta)\cosh(\beta J)-1\}}.
\end{equation}
This measurement corresponds to a simultaneous measurement of the populations of both sites and in Fig.~\ref{fig3N} {\bf (c)} and {\bf (d)} we show its performance. This approach allows us to effectively perform optimal temperature estimation and requires only simple `global' measurements.

\subsection{Implementation}
In the previous section we showed that determining the population of one or both oscillators allowed for a good estimation of the temperature, the precision of which increased with the interaction term $J$. Here we outline a scheme where it is possible to probe the local energy $\big< \hat{a}_1^\dagger \hat{a}_1\big>$, by extending our system to include an ancillary mode. Following~\cite{PhysRevA.91.033631} we consider a single mode with frequency $\omega_C$, free Hamiltonian $\hat{H}_C=\omega_C \left( \hat{c}^\dagger \hat{c} + \frac{1}{2} \right)$, and further assume the mode is pumped according to $\hat{H}_P=i\eta\left( \hat{c}^\dagger - \hat{c} \right)$. We assume the oscillator and the ancillary mode interact according to
\begin{equation}
\hat{H}_I=\kappa \hat{c}^\dagger \hat{c} \hat{a}^\dagger_1 \hat{a}_1.
\end{equation}
We now study the dynamics by solving the master equation 
\begin{equation}
\label{master}
\partial_t \varrho = -i \left[ \hat{H}+\hat{H}_C+\hat{H}_I+\hat{H}_P, \varrho \right] + \frac{\gamma}{2} \left( 2 \hat c \varrho \hat c^\dagger - \hat c^\dagger \hat c \varrho - \varrho \hat c^\dagger \hat c  \right), 
\end{equation}
\begin{figure}[t]
\begin{center}
{\bf (a)}\hskip0.4\columnwidth{\bf (b)}\\
\includegraphics[width=0.45\columnwidth]{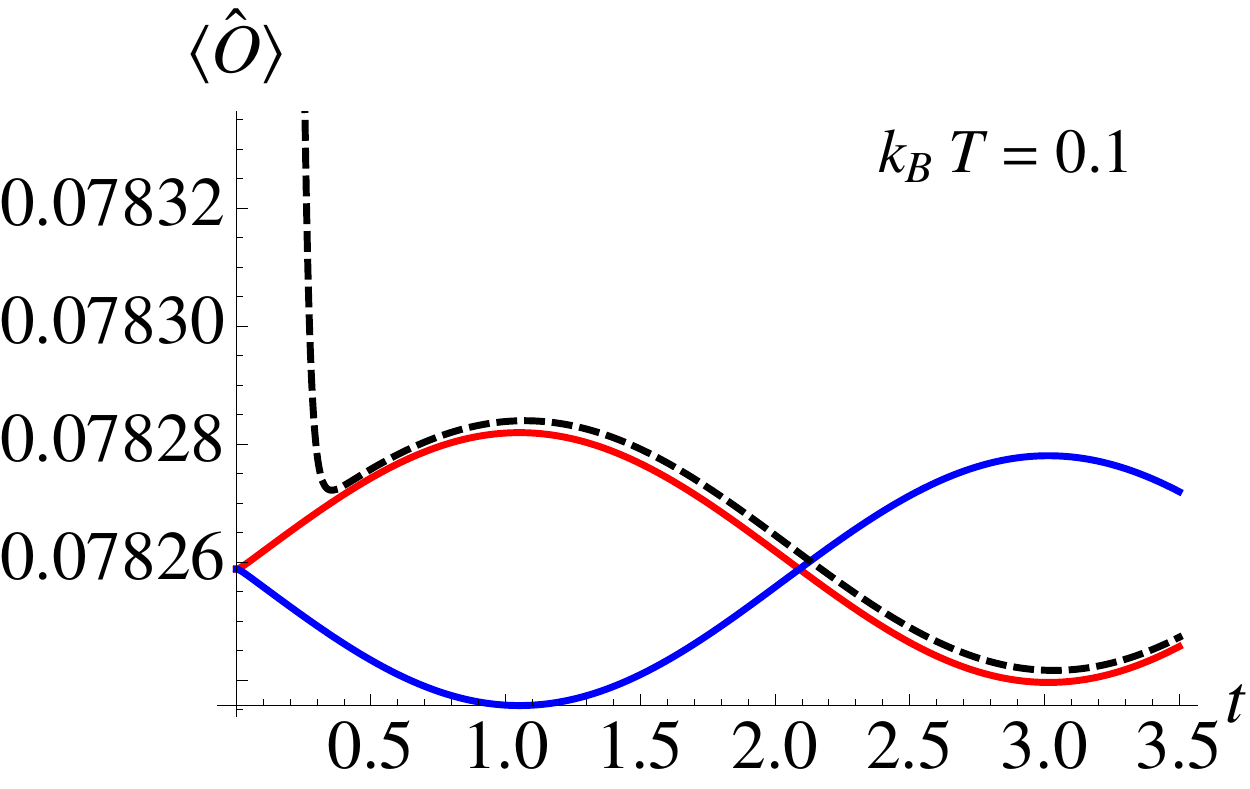}~\includegraphics[width=0.45\columnwidth]{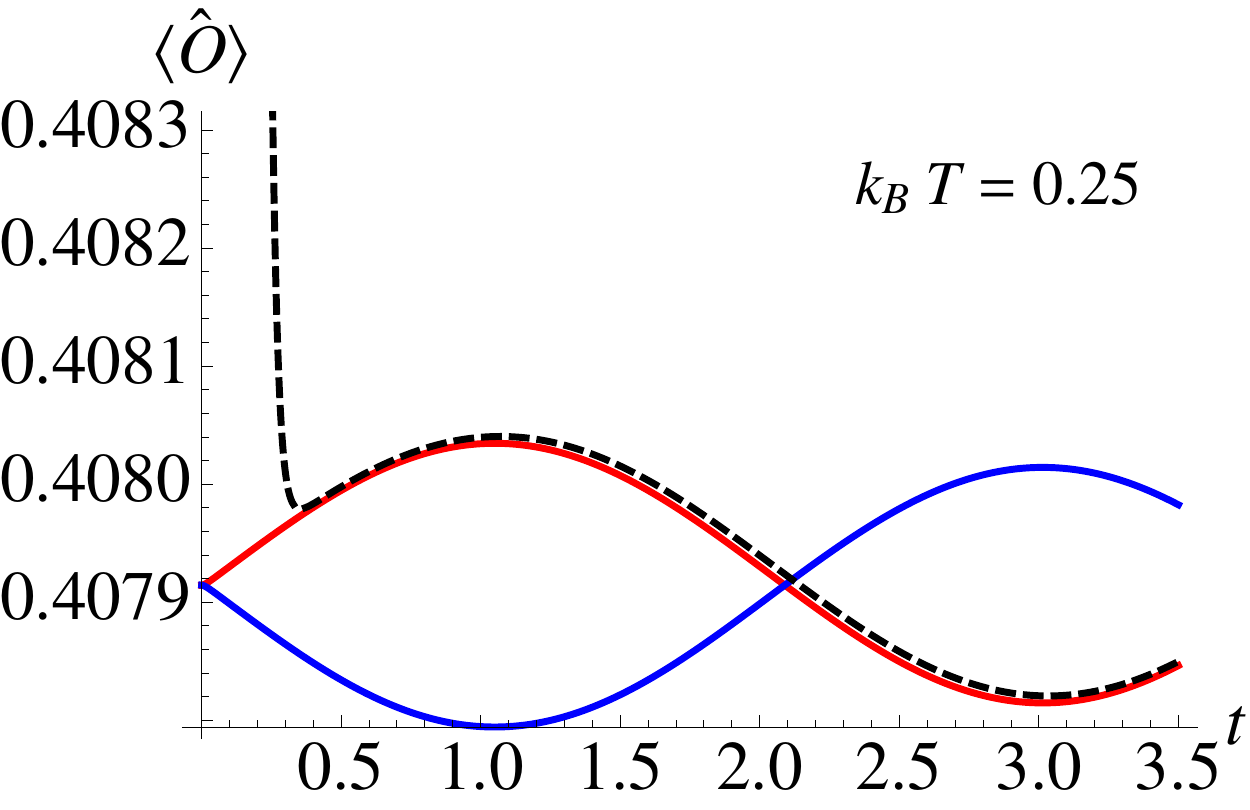}\\
{\bf (c)}\hskip0.4\columnwidth{\bf (d)}\\
\includegraphics[width=0.45\columnwidth]{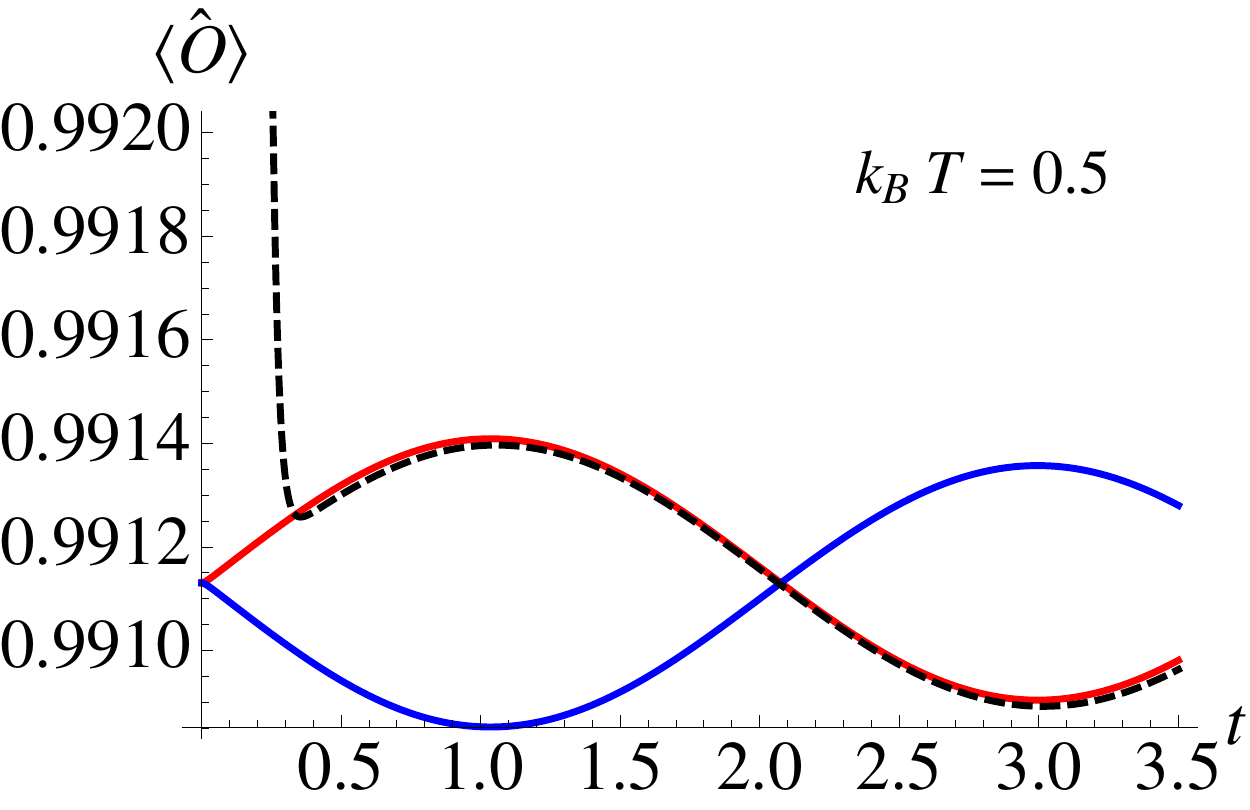}~\includegraphics[width=0.45\columnwidth]{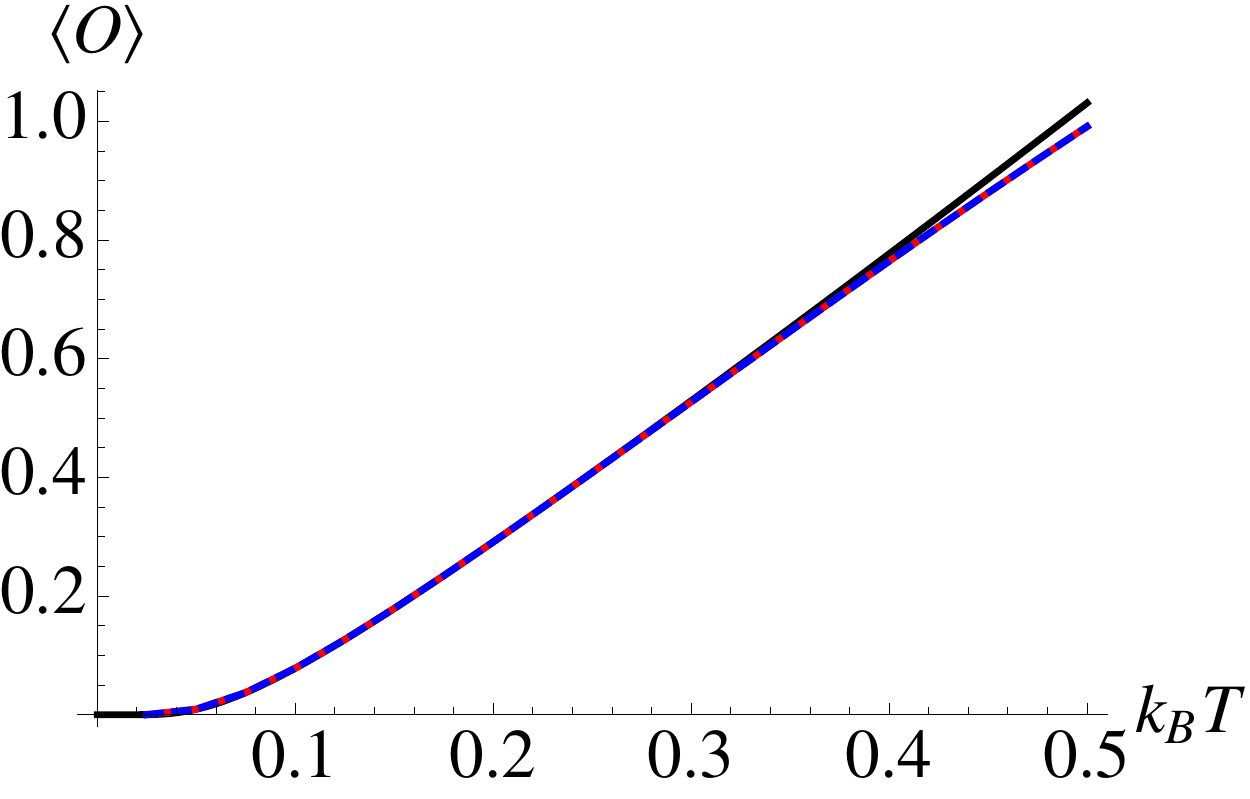}
\end{center}
\caption{Fixing $J=0.8$, $U=0$, $\Delta=0$, we show $\big< \hat{a}_1^\dagger \hat{a}_1 \big>$ (red), $\big< \hat{a}_2^\dagger \hat{a}_2 \big>$ (blue) and $\big< \hat{p}_C \big>$ (black dashed) for {\bf (a)} $k_B T=0.1$, {\bf (b)} $k_B T=0.25$, and {\bf (c)} $k_B T=0.5$ with $\eta=1$, $\gamma=100$, and $\kappa=0.1$. In panel  {\bf (d)} we show the results of our numerical simulations compared to the analytic expression for $\big< \hat{a}_1^\dagger \hat{a}_1 \big>$ (solid black line). The deviation at large temperatures is simply due to numerical truncation.}
\label{fig4N}
\end{figure}
where we have taken the number of thermal photons to be zero~\cite{PhysRevA.91.033631}. We assume the coupled oscillators are initially in their thermal state, Eq.~\eqref{gibbs}, while the ancillary mode is initially in a coherent state $\ket{\alpha}=\sum_n \frac{e^{-\alpha^2/2} \alpha^n}{\sqrt{n!}} \ket{n}$. Following Ref.~\cite{PhysRevA.91.033631} we make some further assumptions, namely that the interaction between the ancillary mode and the oscillator is smaller than the oscillator coupling strength, i.e. $\kappa\! \ll \! J$. In this way we ensure that the additional interaction only weakly affects the dynamics of the system, allowing the ancillary mode to act as a very effective non-destructive probe. From Ref.~\cite{PhysRevA.91.033631}, if we measure $\hat{p}_C=\frac{i}{\sqrt{2}}\left( \hat{c}^\dagger - \hat{c} \right)$, we can immediately determine $\big< \hat{a}_1^\dagger \hat{a}_1 \big>$ according to
\begin{equation}
\big< \hat{a}_1^\dagger \hat{a}_1 \big>=-\frac{\gamma^2}{4\sqrt{2}\kappa\eta}\big< \hat{p}_C \big>.
\end{equation}

In Fig.~\ref{fig4N} we examine the effectiveness of this approach, taking $J=0.8$ we see with a suitably large damping we quickly reach an almost identical expectation value. Furthermore, the interaction is sufficiently weak such that observed expectation values are accurate to $\sim10^{-3}$. This approach could be realised by placing one oscillator inside a single mode leaky cavity. By ensuring the coupling between the cavity field and the oscillator is sufficiently weak all the necessary ingredients outlined above can be achieved. We remark that the same scheme could be used for the loaded double-well potential addressed in Sec.~\ref{DWs} without any modification.

\section{Estimation of the coupling constant and multi-parameter estimation}
It is straightforward to identify the optimal precision bound on estimation of $J$, as well as the measurement strategy that achieves it. In what follows we set $U=0$. To begin with, if the two oscillators are at resonance, the Hamiltonian term conjugate to $J$ commutes with the rest of the Hamiltonian, i.e., $[\hat H,\partial_J \hat H]=0$. On this account, at thermal equilibrium the optimal observable to estimate $J$ is $\hat A=\partial_J \hat H=\hat a^\dagger_1 \hat a_2 + \hat a_1 \hat a^\dagger_2$ \cite{Mehboudi_in_preparation}. The corresponding sensitivity, which is equivalent to the QFI, is simply proportional to $\partial_J\big<\hat A\big>$,
\begin{equation}
\label{tunnellingQFI}
{\cal H}_J = \frac{\beta ^2 \left\{\cosh (\beta ) \cosh (\beta  J)-1 \right \}}{\left\{ \cosh (\beta )-\cosh (\beta J)\right\}^2}.
\end{equation}
Interestingly, since $[\hat H,\hat A]=0$, it is possible to simultaneously estimate $J$ and $T$ with optimal precision. However, this is achieved only by performing non-trivial global measurements.
\begin{figure*}
	\centering
	\includegraphics[width=0.6\textwidth]{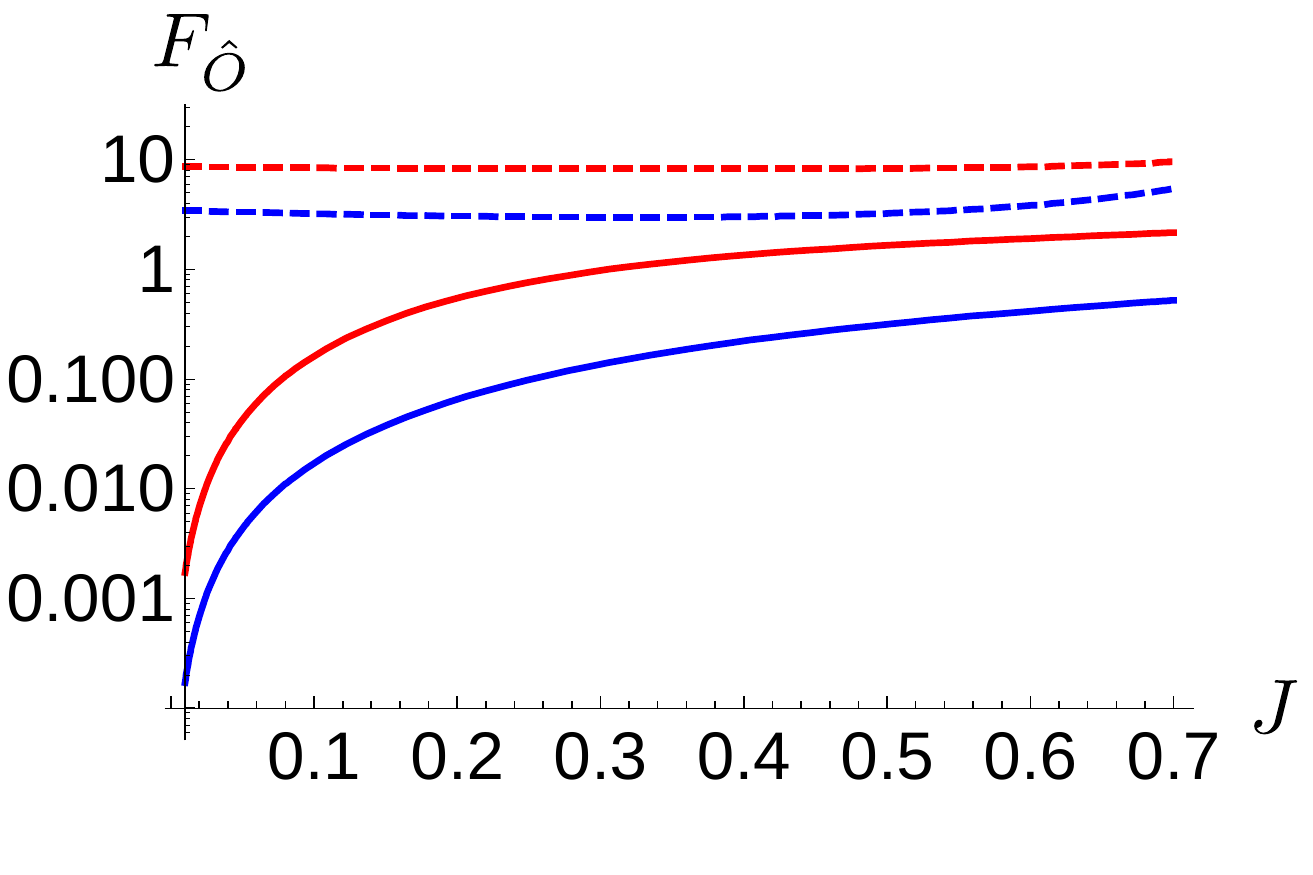}		
	\caption[fig]{(Solid) The FI achieved by measuring $\hat O=n_1$, versus $J$. We consider two different temperatures: $k_BT=0.25$ (red) and $k_BT=0.5$ (blue). For comparison, we benchmark this FI against the QFI (dashed lines), which might be achieved by measuring $\hat O=\hat A$. Note that, unlike thermometry, the local measurements fail to compete with the optimal one. Also note that increasing the temperature reduces the precision of both local and global estimations.}
	\label{fig5N}
\end{figure*}
%--------------------------------------

Since the interaction coupling is coupled to a non-local operator, naturally we do not expect it to be detected optimally via local measurements. We can confirm this conjecture by exploring the FI of such local measurements. In particular, we examine the sensitivity of the local number operator, i.e., $F_{n_1}=\left|\partial_J\big<a_{_1}^{\dagger}a_{_1}\big>\right|^2/\mathrm{Var}(a_{_1}^{\dagger}a_{_1})$. As depicted in Figure~\ref{fig5N}, the local estimations are not very efficient for the estimation of $J$, and the optimal measurement notably outperforms them. Moreover, it can be seen that by decreasing the temperature, we can estimate $J$ with a higher precision.

Finally, for a non resonant scenario with $\Delta\neq 0$, the optimal observable is found to be exactly the same as the resonant case, i.e., $\hat A$ \cite{Mehboudi2017,monras2013phase,PhysRevA.89.032128}. However, the optimal sensitivity is different from Eq.~\eqref{tunnellingQFI}, and is given by ${\cal H}_J^{\Delta}=\beta C(\Delta)^2\partial_J\big<\hat A\big>$, where the extra coefficient is
\begin{equation}
	C(\Delta)=\frac{\tanh\Delta/2T}{\Delta/2T},
\end{equation}
and is arising from the detuning. This coefficient is a monotonically decreasing function of $\Delta$, and $0\leq C(\Delta)\leq 1$ with the upper bound holding at resonance. Therefore, the detuning leads to a reduction of precision, similar to thermometry. In addition, another influence of the detuning is that, the optimal observable $\hat A$, does not commute with the system Hamiltonian anymore, hence optimal estimation of $T$ and $J$ is not possible simultaneously.
%--------------------------------------
\section{Conclusions}
\label{conclusion}
We have examined the ultimate limits on temperature estimation for a widely applicable model, two interacting bosonic modes held in a common bath. Interestingly we find the mechanism with which the system interacts with its bath, namely if this is with or without particle exchange, leads to very different results. If the total particle number is fixed, thus effectively modelling a loaded double well potential, we find that increasing the tunnelling rate between the two wells reduces the ability to accurately estimate the temperature. Additionally, strong self interactions are shown to quite severely affect the estimation of temperature. Conversely, when particle exchange with the bath is permitted, increased coupling between the modes (which in this context is closely related to the tunnelling) actually increases the thermometry ability. Furthermore, the effect that the non-linear term (which is the complement to the self-interactions) has only a weak effect in the small temperature range. In both settings ensuring both modes are on resonance is shown to be crucial. We have shown that all these results can be succinctly explained by examining the effect changing the Hamiltonian parameters have on the energy spectra. Finally, we have assessed experimentally implementable schemes to estimate the temperature, showing that near optimal (Heisenberg limited) estimation can be achieved using only local energy measurements. We have presented a simple scheme that requires only weakly coupling an ancillary mode to our system, such that it acts as a non-destructive probe to achieve the required energy measurements. We remark that although our analysis has focused on thermometry, the general approach presented can be considered for virtually any other parameter estimation scheme. 

\section*{Acknowledgements}
This work is supported by the EU Collaborative Project TherMiQ (Grant Agreement 618074), the Spanish MINECO Projects No. FIS2013-40627-P and No. FIS2016- 86681-P (AEI/FEDER,UE), the Generalitat de Catalunya CIRIT (2014-SGR-966), the Julian Schwinger Foundation (grant JSF-14-7-0000), the Royal Society Newton Mobility Grant (grant NI160057), and the DfE-SFI Investigator Programme (grant 15/IA/2864). Part of this work was supported by COST Action MP1209 Thermodynamics in the quantum regime.
 
\section*{References}
\bibliographystyle{iopart-num}
\bibliography{arXiv_bib_noURL}
\end{document}